\documentclass[12pt]{article}
\pdfoutput=1
\usepackage{subfigure}
\usepackage{amssymb,amsmath}
\usepackage{graphicx}
\usepackage{color}
\usepackage[colorlinks=true
,urlcolor=blue
,citecolor=blue
,linkcolor=blue
,pagecolor=blue
,linktocpage=true
,pdfproducer=medialab
]{hyperref}
\usepackage[a4paper,width=15.2cm]{geometry}
\makeatletter \renewcommand{\@dotsep}{10000} \makeatother
\def\be{\begin{equation}}
\def\ee{\end{equation}}
\def\bea{\begin{eqnarray}}
\def\eea{\end{eqnarray}}
\def\bi{\begin{itemize}}
\def\ei{\end{itemize}}

\def\to{\rightarrow}

\def\tg{\tilde g}

\def\tq{\tilde q}

\def\tw{\tilde\chi^{\pm}}
\def\twm{\tilde\chi^{-}}
\def\tz{\tilde\chi^0}
\def\alt{\lesssim}
\def\agt{\gtrsim}
\def\R{R_{tb\tau}}

\newcommand\prd[3]{{\it Phys.\ Rev.\ }{\bf D #1} (#2) #3}
\newcommand\prl[3]{{\it Phys.\ Rev.\ Lett.\ }{\bf #1} (#2) #3}
\newcommand\npb[3]{{\it Nucl.\ Phys.\ }{\bf B #1} (#2) #3}
\newcommand\plb[3]{{\it Phys.\ Lett.\ }{\bf B #1} (#2) #3}

\newcommand\jhep[3]{{\it J. High Energy Phys.\ }{\bf #1} (#2) #3}

\def\tg{\tilde g}
\def\tw{\widetilde \chi^{\pm}}
\def\tz{\widetilde \chi^0}

\def\tq{\tilde q}

\newcommand{\beq}{\begin{equation}}
\newcommand{\eeq}{\end{equation}}

\begin{document}
%Remove date before submitting to arXiv
\date{\today}

\begin{center}

 {\Large\bf A heavier gluino from $t$-$b$-$\tau$ Yukawa-unified SUSY  
 } \vspace{1cm}
\end{center}

\begin{center}
{\Large
Howard Baer$^{a,}$\footnote{
Email: baer@nhn.ou.edu},
Shabbar Raza $^{b,}$\footnote{
Email: shabbar@udel.edu. On study leave from:
Department of Physics, FUUAST, Islamabad, Pakistan.}
and
Qaisar Shafi $^{b,}$\footnote{
Email: shafi@bartol.udel.edu. }
}

\vspace{0.75cm}

{\it $^a$
Dep't of Physics and Astronomy,
University of Oklahoma, \\ Norman, OK 73019, USA
}\\
{\it  $^b$
Bartol Research Institute, Department of Physics and Astronomy,\\
University of Delaware, Newark, DE 19716, USA %\\ \vspace{2mm}
}

\vspace{1.5cm}
\section*{Abstract}
\end{center}
Supersymmetric models with $t-b-\tau$ Yukawa coupling unification and unified gaugino masses 
at the GUT scale-- with $\mu >0$-- show a mild preference for light gluino masses $m_{\tg}\alt$ 500 GeV.  
This range of $m_{\tg}$ is now essentially ruled out by LHC searches.
We show that a heavier gluino with $m_{\tg}\sim 0.5-3\,{\rm TeV}$ can also be compatible with 
excellent $t-b-\tau$ Yukawa coupling unification in supersymmetric models with 
non-universal Higgs masses (NUHM2). 
The gluino in such models is the lightest colored sparticle, while the squark sector 
displays an inverted mass hierarchy with $m_{\tq}\sim 5-20$ TeV.
We present some LHC testable benchmark points for which the lightest 
Higgs boson mass $m_h\simeq 125$ GeV. We also discuss LHC signatures of Yukawa-unified models
with heavier gluinos. We expect gluino pair production followed by decay to final states
containing four $b$-jets plus four $W$-bosons plus missing $E_T$ to occur at possibly
observable rates at LHC.
\newpage

%%%%%%%%%%%%%%%%%%%%%%%%%%%%%%%%%%%%%%%%%%%%%%%%%%%%%%%%%%%%
\renewcommand{\thefootnote}{\arabic{footnote}}
\setcounter{footnote}{0}

%%%%%%%%%%%%%%%%%%%%%%%%%%%%%%%%%%%%%%%%%%%%%%%%%%%%%%%%%%%%%

%\baselineskip 36pt
% Main body
%%%%%%%%%%%%%%%%%%%%%%%%%%
%\baselineskip 18pt
%%%%%%%%%%%%%%%%%%%%%%%%%%

\section{Introduction}
\label{sec:intro}

Unification at $M_{GUT}$ ($\sim 2\times 10^{16}$ GeV) of $t-b-\tau$ Yukawa couplings\cite{old,Anderson} 
is largely inspired by the simplest supersymmetric (SUSY) $SO(10)$ or $SU(4)_c \times SU(2)_L \times SU(2)_R$ 
models\cite{pati}. 
It has become clear in recent years\cite{hb,bf,abbbft,bdr,alt,bartol1,bartol2,wagner} that imposing 
$t-b-\tau$ Yukawa coupling unification has important consequences for the sparticle and 
higgs mass spectrum of the minimal supersymmetric standard model (MSSM). 
The successful launch of the Large Hadron Collider (LHC) has provided important new impetus 
for these studies\cite{yu_lhc7,yu_lhc14,yu_atlas}. 
For analogous discussion of $b-\tau$ unification, see Ref.~\cite{btau}.

The parameter space of $SO(10)$ SUSY GUT models for this investigation is given by
\be
m_{16},\ m^2_{H_u},\ m^2_{H_d}  ,\ m_{1/2},\ A_0,\ \tan\beta,\ sign(\mu ),
\ee
where $m_{16}$ is the unified matter scalar mass, $m^2_{H_u}$ and $m^2_{H_d}$ are the GUT scale Higgs soft masses,
$m_{1/2}$ is the unified gaugino mass, $A_0$ is the coefficient of the soft supersymmetry breaking (SSB) 
trilinear term, and $\tan\beta$ is the ratio of Higgs field vevs.
In order to allow for an appropriate radiative breaking of electroweak symmetry, the two GUT scale 
Higgs doublet masses must be split\cite{mur} according to $m_{H_u}^2<m_{H_d}^2$. This splitting 
might arise due to $SO(10)$ $D$-terms (D-term splitting) 
or via GUT-scale threshold corrections\cite{bdr} (the Higgs splitting, or $HS$ model). 
With the MSSM superpotential parameter $\mu >0$, this scenario predicts an inverted scalar mass 
hierarchy (IMH)\cite{imh} in the squark sector, 
wherein third generation squarks have masses in the few TeV range, 
while the first two generations of squarks have mass in the 5-30 TeV range\cite{abbbft}. 
The IMH allows one to reconcile a decoupling solution to the SUSY flavor and $CP$ problems with 
relatively low fine-tuning in the EWSB sector.
A successful implementation of the IMH scheme requires GUT-scale SSB terms to be related
as $A_0^2\simeq 2m_{10}^2\simeq 4m_{16}^2$, with unified third generation Yukawa couplings
and the unified gaugino mass $m_{1/2}$ on the low side: typically sub-TeV. 

One particularly important prediction concerns the gluino which turns out to be the 
lightest colored SUSY particle.\footnote{These predictions are obtained under the assumption 
that the lightest neutralino is the lightest MSSM particle (LMP). 
This class of Yukawa-unified models tends to predict a thermal neutralino 
relic abundance $\Omega_{LMP} h^2 >> 1$ so that the neutralino LMP in this scenario is 
not a viable dark matter candidate. 
To overcome this drawback, one proposal\cite{bs,bkss} is to invoke axion physics 
and arrange for the lightest neutralino to decay before nucleosynthesis into an axino, which now plays 
the role of lightest SUSY particle (LSP). 
A combination of axions and axinos could then make up the dark matter content of the universe.
}
Since $m_{1/2}$ is favored to be small in comparison to $m_{16}$, there is some tendency
in $t-b-\tau$ unified models for $m_{\tg}\alt 500$ GeV, which should be within range of 
SUSY searches at LHC operating with $\sqrt{s}=7$ TeV (LHC7)\cite{lhc7}.
In this case, due to the large $b$-quark Yukawa coupling and the inverted squark mass
spectrum, gluinos are expected to dominantly decay by 3-body modes such as $\tg\to b\bar{b}\tz_i$, 
leading to final states at LHC consisting of multiple $b$-jets $+MET$\cite{yu_lhc7,yu_lhc14}.

In fact, recent searches by the ATLAS experiment with less than 1 fb$^{-1}$ of data already
exclude $m_{\tg}\alt 500$ GeV by searching for multijet plus missing $E_T$ (MET) plus
one or more tagged $b$-jets\cite{yu_atlas}. Also, direct searches for gluinos and squarks
under the assumption of unified gaugino masses 
by Atlas and CMS (again with $\sim 1$ fb$^{-1}$ of data) typically exclude
$m_{\tg}<550-750$ GeV (depending on search techniques)\cite{atlas,cms}.
Based on this critical input from experiment, the question arises: 
Are $t-b-\tau$ unified models now excluded by LHC searches, or can Yukawa-unified 
solution be found with heavier gluinos with mass beyond current LHC reach?
And if such solutions are found, what is the nature of the SUSY signal which is expected in
near future runs of LHC7?

Our main goal in this paper is to determine whether Yukawa-unified solutions with a heavier gluino
can exist. In fact, we find numerous solutions, some of which are presented as 
a new set of benchmark points with $m_{\tg}$ in the range $0.5-3$ TeV. 
In these solutions, the gluino retains its position as the lightest colored SUSY particle,
while squarks remain in the multi-TeV range.
Thus, we expect in this class of models that LHC searches should focus on gluino pair
production. However, for these heavier gluino solutions, the $\tg$ is expected to decay via
$\tg\to tb\tw_1$ or $t\bar{t}\tz_i$. After $t\to bW$ and $\tw_1\to \tz_1 W$ decays, we
expect gluino pair production final states to contain typically four $b$-jets, four $W$
bosons plus MET. These are in rather sharp contrast with models containing $m_{\tg}\alt 500$
GeV, where multi-$b$-jets$+MET$ final states are expected, but without the numerous 
on-shell $W$ bosons. 

Note that due to potential threshold corrections which could arise from a variety of 
sources including a more complicated Higgs sector, higher order interaction terms, etc., 
we do not insist on exact (or perfect) unification of the three Yukawa couplings. 
Instead, in this paper Yukawa unification realized at the 10\% level (or better) is 
considered to yield an acceptable scenario. 
In practice, we find solutions with heavy ($\sim 2-3$ TeV) gluino masses that are 
associated with Yukawa unification at a few percent level. 
Somewhat lighter gluino masses (~ 1-1.5 TeV) are accompanied by essentially perfect Yukawa unification!
As expected, the squark masses display an inverted mass hierarchy, 
with the lightest (third family) squark masses ranging between 1 to 10 TeV. 
The first two family squarks turn out to be considerably heavier, of order 8-28 TeV.
In the benchmark points that we highlight in this paper, 
the mass of the SM-like Higgs boson is of order 124-126 GeV, 
a value which is consistent with results from recent ATLAS and CMS Higgs searches\cite{higgslhc}.

\section{Phenomenological constraints and scanning procedure}
\label{sec:scan}

We employ the ISAJET~7.80~\cite{isajet} package Isasugra\cite{isasugra}  to perform random
scans over the fundamental parameter space. 
In this package, the weak scale values of gauge and third generation Yukawa
couplings are evolved to $M_{\rm G}$ via the MSSM renormalization
group equations (RGEs) in the $\overline{DR}$ regularization scheme.
We do not strictly enforce the unification condition $g_3=g_1=g_2$ at $M_{\rm G}$, 
since a few percent deviation from unification can be assigned to unknown GUT-scale threshold
corrections~\cite{Hisano:1992jj}.
The deviation between $g_1=g_2$ and $g_3$ at $M_{G}$ is no
worse than $3-4\%$.
For simplicity  we do not include the Dirac neutrino Yukawa coupling
in the RGEs, whose contribution is usually small\cite{dr3}.

The various HS model boundary conditions are imposed at $M_{\rm G}$ and all the SSB
parameters, along with the gauge and Yukawa couplings, are evolved
back to the weak scale $M_{\rm Z}$.
In the evaluation of Yukawa couplings, the SUSY threshold corrections~\cite{pbmz} 
are taken into account at the common scale $M_{\rm SUSY}= \sqrt{m_{{\tilde t}_L}m_{{\tilde t}_R}}$. 
The entire parameter set is iteratively run between $M_{\rm Z}$ and $M_{\rm G}$ using the 
full 2-loop RGEs until a stable solution is obtained. 
To better account for leading-log corrections, one-loop step-beta functions are adopted 
for gauge and Yukawa couplings, and the SSB parameters $m_i$ are extracted from 
RGEs at multiple scales $m_i=m_i(m_i)$. 
The RGE-improved 1-loop effective potential is minimized at $M_{\rm SUSY}$, which effectively
accounts for the leading 2-loop corrections. Full 1-loop radiative
corrections are incorporated for all sparticle masses.

The requirement of  radiative electroweak symmetry breaking (REWSB) imposes an important theoretical
constraint on the parameter space.
In order to reconcile REWSB with Yukawa unification, the MSSM Higgs
 soft supersymmetry breaking (SSB) masses should be split in
such way  that $m^2_{H_{d}}/ m^2_{H_u}> 1.2$  at  $M_{\rm G}$ \cite{Olechowski:1994gm}.
As mentioned above, the MSSM doublets reside in the 10 dimensional
representation of $SO(10)$ GUT  for Yukawa unification condition to hold. 
In the gravity mediated  supersymmetry breaking  scenario\cite{Chamseddine:1982jx}, 
the required splitting in the Higgs sector can be generated by involving 
additional Higgs fields\cite{bartol2}, or via D-term contributions\cite{Drees:1986vd}.
In our Yukawa-unified SUSY spectrum calculations, the lightest neutralino is
always turns out to be the LMP.

We have performed Markov-chain Monte Carlo (MCMC) scans for the following parameter range:
\begin{align}
0\leq  m_{16}  \leq 30\, \rm{TeV} \nonumber \\
0\leq   m_{H_u} \leq 35\, \rm{TeV} \nonumber \\
0\leq    m_{H_d} \leq 35\, \rm{TeV} \nonumber \\
0 \leq m_{1/2}  \leq 5 \, \rm{TeV} \nonumber \\
30\leq \tan\beta \leq 60 \nonumber \\
-3\leq A_{0}/m_0 \leq 3
 \label{parameterRange}
\end{align}
with  $\mu > 0$ and  $m_t = 173.3\, {\rm GeV}$  \cite{:2009ec}.
Note that our results are not
too sensitive to one or two sigma variation in the value of $m_t$  \cite{bartol2}.
We use $m_b^{\overline{DR}}(m_Z)=2.83$ GeV which is hard-coded into ISAJET.

In scanning the parameter space, we employ the Metropolis-Hastings
algorithm as described in \cite{Belanger:2009ti}. 
The data points collected all satisfy the requirement of REWSB,
with the neutralino in each case being the LMP. After collecting the data, we impose
the mass bounds on all the particles\cite{Nakamura:2010zzi} and use the 
IsaTools package~\cite{bsg,bmm} and Ref. \cite{mamoudi}
to implement the various phenomenological constraints. 
We successively apply the following experimental constraints on the data that
we acquire from Isasugra:
\begin{table}[h]\centering
\begin{tabular}{rlc}
$m_h~{\rm (lightest~Higgs~mass)} $&$ \geq\, 114.4~{\rm GeV}$          &  \cite{Schael:2006cr} \\
%$m_{\tilde g}  $&$ \geq \, 800~{\rm GeV}$ \\
$BR(B_s \rightarrow \mu^+ \mu^-) $&$ < \, 1.1 \times 10^{-8}$        &   \cite{CMS_plus_LHCb}      \\
$2.85 \times 10^{-4} \leq BR(b \rightarrow s \gamma) $&$ \leq\, 4.24 \times 10^{-4} \;
 (2\sigma)$ &   \cite{Barberio:2007cr}  \\
$0.15 \leq \frac{BR(B_u\rightarrow
\tau \nu_{\tau})_{\rm MSSM}}{BR(B_u\rightarrow \tau \nu_{\tau})_{\rm SM}}$&$ \leq\, 2.41 \;
(3\sigma)$ &   \cite{Barberio:2008fa}  \\
%$\Omega_{\rm CDM}h^2 $&$ =\, 0.111^{+0.028}_{-0.037} \;(5\sigma)$ &
%\cite{Komatsu:2008hk} \\ 
%$ 0 \leq \Delta(g-2)_{\mu}/2 $ & $ \leq 55.6 \times 10^{-10} $ & \cite{Bennett:2006fi}
\end{tabular}\label{table}
\end{table}

As far as the muon anomalous magnetic moment $a_{\mu}$ is concerned, we require that the benchmark
points are at least as consistent with the data as the Standard Model is. For a presentation of $(g-2)_{\mu}$ 
values in NUHM2 models, see \cite{h125}.

\section{A heavier gluino from Yukawa-unified SUSY}
\label{sec:results}

\begin{figure}[htp!]
\centering
\subfiguretopcaptrue

\subfigure{
\includegraphics[totalheight=5.5cm,width=7.cm]{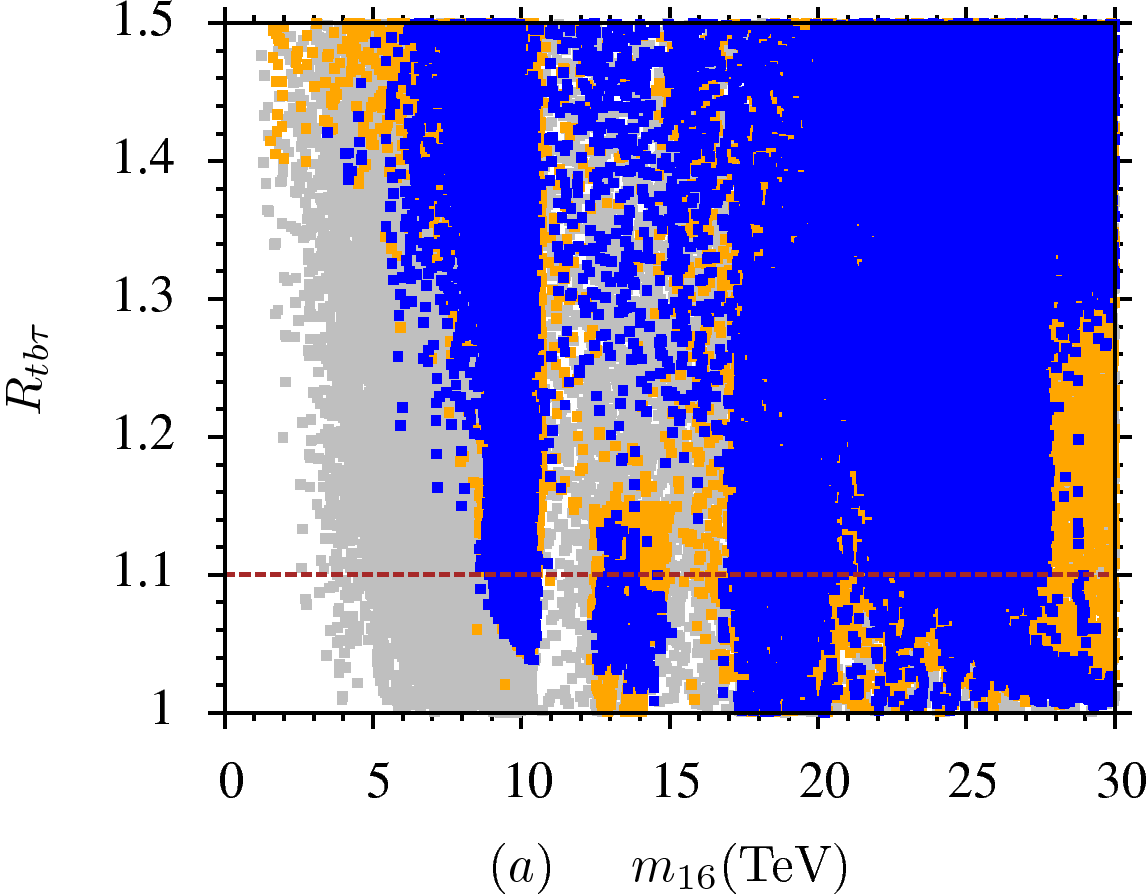}
}
\subfigure{
\includegraphics[totalheight=5.5cm,width=7.cm]{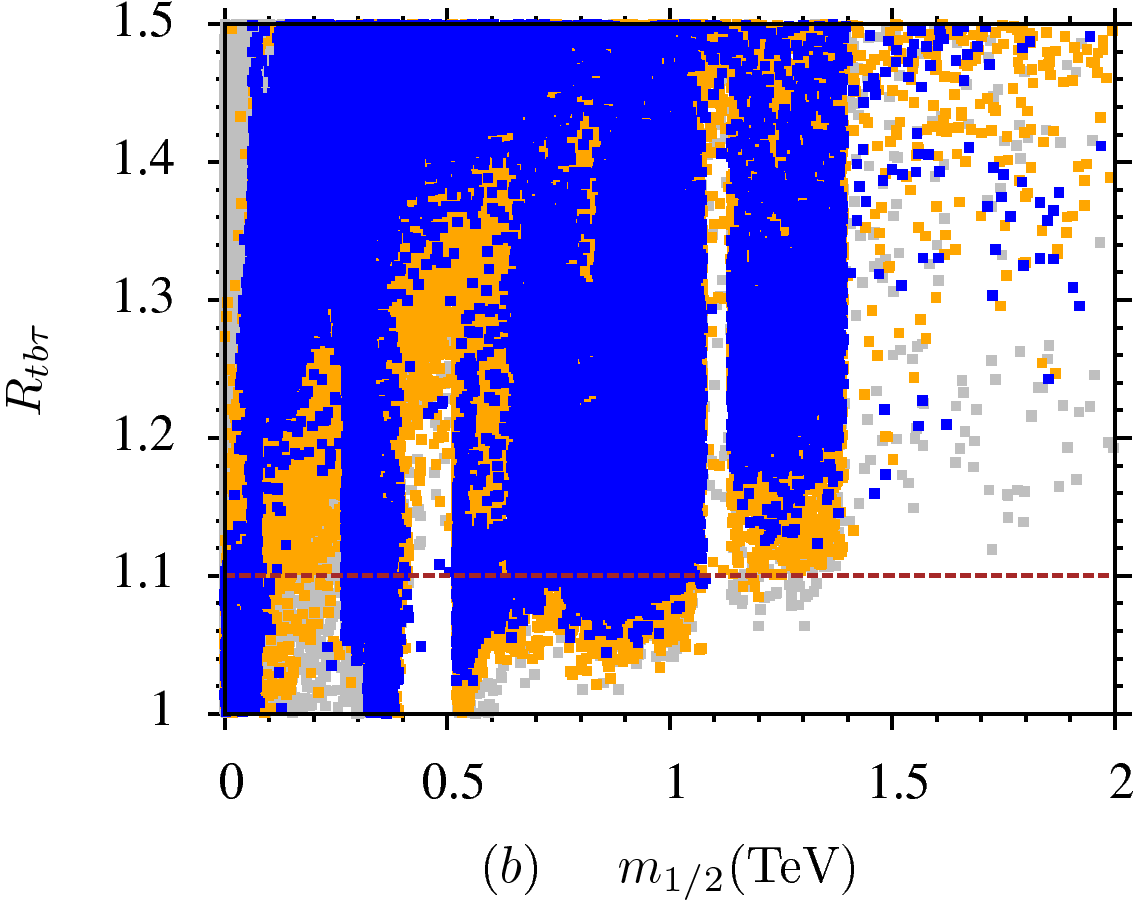}
}
\subfigure{
\includegraphics[totalheight=5.5cm,width=7.cm]{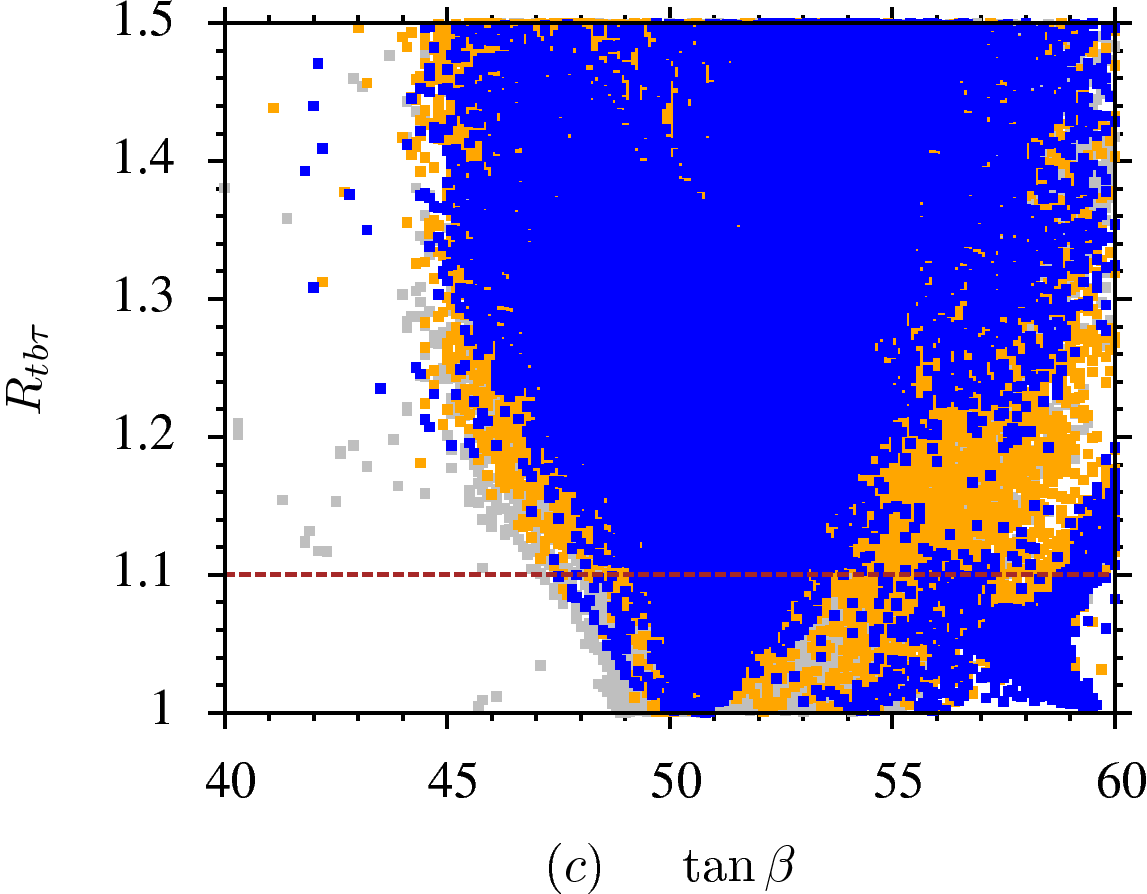}
}
\subfigure{
\includegraphics[totalheight=5.5cm,width=7.cm]{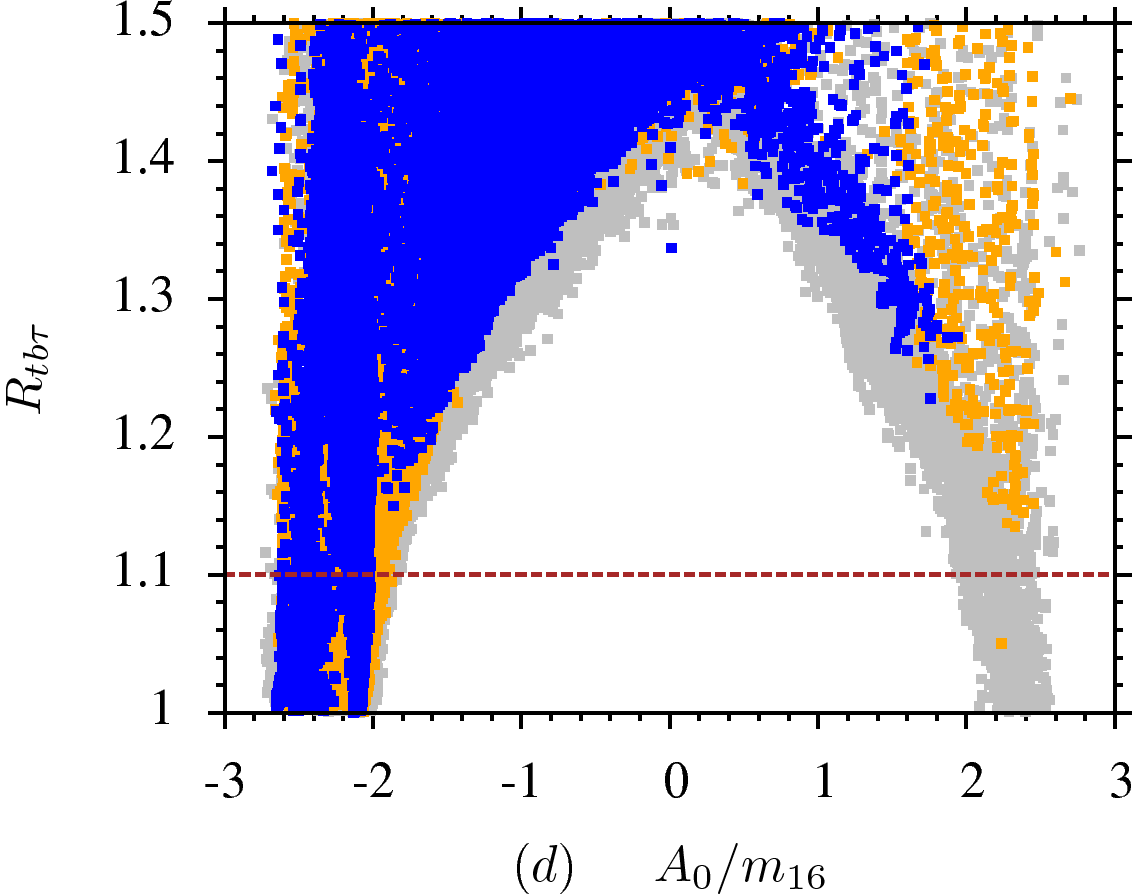}
}
\subfigure{
\includegraphics[totalheight=5.5cm,width=7.cm]{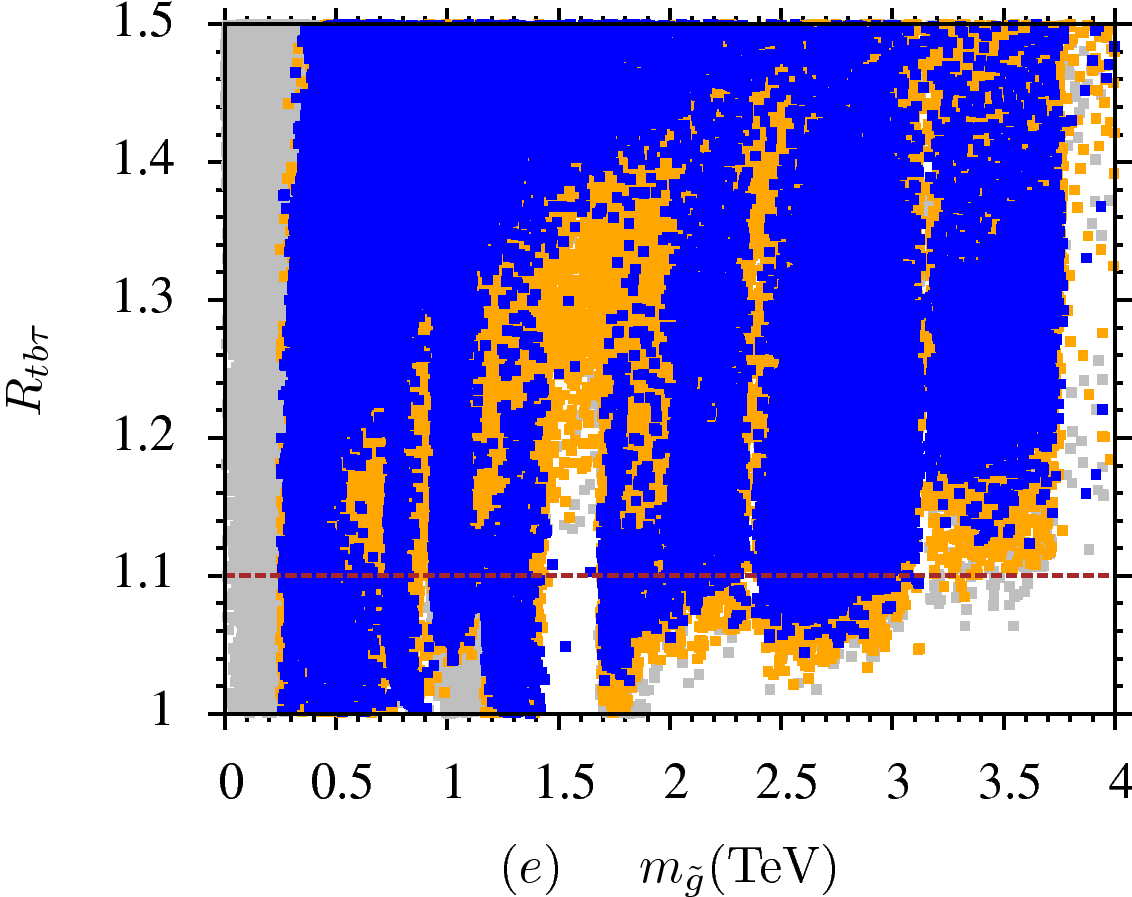}
}

\caption{Plots in $m_{16}-R_{tb\tau}$, $m_{1/2}-R_{tb\tau}$,
$\tan\beta-R_{tb\tau}$, $A_{0}/m_{16}-R_{tb\tau}$, and
$m_{\tilde g}-R_{tb\tau}$ planes.
Gray points are consistent with REWSB and neutralino LSP. Orange points satisfy mass bounds (including $ m_h$ in the range $115-131 \,{\rm GeV }$ and $m_{\tilde g}\ge 0.5\,{\rm TeV}$), constraints from $BR(B_s\rightarrow \mu^+ \mu^-)$,  $BR(B_u\rightarrow \tau \nu_{\tau})$  and $BR(b\rightarrow s \gamma)$.
Blue point solutions belong to a subset of orange points and represent $m_h$ in the range $123-127 \,{\rm GeV }$.
}
\label{fig:param}
\end{figure}
%%%%%%%%%%%

In order to quantify Yukawa coupling unification, we define the quantity $R_{tb\tau}$ as
\begin{equation}
R_{tb\tau}=\frac{ {\rm max}(y_t,y_b,y_{\tau})} { {\rm min} (y_t,y_b,y_{\tau})}.
\label{eq:R}
\end{equation}
In Fig.~\ref{fig:param} we plot $R_{tb\tau}$ versus the various $SO(10)$ model input parameters.
Gray points are consistent with REWSB and neutralino LSP. 
Orange points satisfy the mass bounds (including $ m_h$ in the range $115-131 \,{\rm GeV }$ and
 and $m_{\tilde g}\ge 0.5\,{\rm TeV}$), 
constraints from $BR(B_s\rightarrow \mu^+ \mu^-)$,  $BR(B_u\rightarrow \tau \nu_{\tau})$  and $BR(b\rightarrow s \gamma)$.
Blue point solutions belong to a subset of orange points and represent $m_h$ in the range $123-127 \,{\rm GeV }$.
In Fig. \ref{fig:param}{\it a}), we see, as is well known, that $m_{16}\agt 10$ TeV for solutions with $R_{tb\tau}<1.1$, 
as required by the inverted scalar mass hierarchy. 
In Fig. \ref{fig:param}{\it b}), we also see that low values of $m_{1/2}$ are favored. 
While previous works favored $m_{1/2}\alt 0.1-0.2$ TeV, here
our dedicated MCMC scans show $t-b-\tau$ unified solutions can also occur for $m_{1/2} $ values in the $0.3-1$ TeV range.
Fig. \ref{fig:param}{\it c}) shows that $\tan\beta\sim 50-60$ is required, while Fig. \ref{fig:param}{\it d})
shows that $A_0\sim -2m_{16}$ is required for the IMH. Our key result here occurs in Fig. \ref{fig:param}{\it e}),
where we plot the value of $m_{\tg}\ vs.\ \R$. Here, we find that while many solutions occur with $m_{\tg}\alt 0.5$ TeV, 
there also exist solutions with near perfect Yukawa unification with substantially heavier gluino masses ranging up to
$m_{\tg}\sim 1.4$ TeV. And if we only require $\R\alt 1.1$, then some solutions can occur with $m_{\tg}$
as large as 3 TeV!

\section{Heavier gluino benchmark points and implications for SUSY searches at LHC}
\label{sec:lhc}

In Table \ref{tab:bm}, we list four benchmark (BM) Yukawa-unified solutions from
Isajet 7.80 with $m_{\tg}>500$ GeV.  Each BM point also has $m_h=125\pm 2$ GeV, so 
all are consistent with the Atlas/CMS hint of a Higgs signal around 125 GeV.

For point 1, with $m_{16}\simeq 21$ TeV, all the squarks and sleptons are far beyond the reach
of LHC. However, for this point, $m_{\tg}=750$ GeV, and so gluinos would be pair-produced at LHC7 
with a cross section of $\sim 60$ fb\cite{yu_lhc7}. In Ref.~\cite{yu_lhc7}, LHC search strategies assumed a much lighter
gluino of mass $\sim 0.3-0.6$ TeV, in which case gluino three body decays to $b\bar{b}\tz_i$ are dominant, 
and the search strategy was to look for collider events containing multiple $b$-jets $+MET$. 
For point 1, at the bottom of the Table we list the dominant gluino branching fractions.
In this case, with $m_{\tg}\sim 750$ GeV, the decay modes $\tg\to t\bar{t}\tz_i$ and $\tg\to t\bar{b}\twm_j$ occur at
substantial rates: in this case $\sim 60\%$. Here, $\twm_1\to \tz_1 W$ at 100\% branching fraction, 
while $t\to bW$ also at 100\%. Thus, gluino pair production for Yukawa-unified benchmarks and a heavier 
gluino lead to final states including four $b$-jets, four on-shell $W$-bosons $+MET$. Since the $W$s
decay into hard isolated leptons over 20\% of the time, these gluino pair production events will
contain {\it high multiplicities of isolated leptons}, including same sign (SS) and opposite-sign (OS) pairs,
trileptons and four-leptons! There are few SM background (BG) processes that can lead to
events containing for instance four $b$-jets plus four isolated leptons. The major BG process would
likely be four top production: $pp\to t\bar{t}t\bar{t}X$. 
%%%%%%%%%%%%%%%%%%%%%%%%%%%%%%%%%%%%%%%%

\begin{table}[h!]
\label{table1}
\centering
\begin{tabular}{lcccc}
\hline
\hline
                 & Point 1 & Point 2 & Point 3 & Point 4 \\
\hline
$m_{16}$         & 21370 & 20230  & 18640  & 26130 \\
$m_{1/2} $       & 93.41   & 364    & 579    & 1021 \\
$A_0/m_{16}$            & -2.43   & -2.13  & -2.09  & -2.11 \\
$\tan\beta$      & 57.2    & 51     & 50     & 52 \\
$m_{H_{d}}$      & 22500.0 & 26770  & 24430  & 34210 \\
$m_{H_{u}}$      & 13310.0  & 23260  & 21780  &30590  \\
\hline
$m_h$            & 126.7   &125     & 124 &124 \\
$m_H$            & 9389 & 3192   & 3145&4066\\
$m_A$            & 9328 & 3171   & 3125 &4040 \\
$m_{H^{\pm}}$    & 9390 & 3193   & 3147 &4067\\

\hline
$m_{\tilde{g}}$  & 750     & 1375   &    1853 & 2991      \\
$m_{\tilde{\chi}^0_{1,2}}$
                 &122, 285 & 232, 491  &323,661 &557,1114 \\
$m_{\tilde{\chi}^0_{3,4}}$
                 &19295, 19295 & 6048,6048  &4570,4571 &6315,6315 \\

$m_{\tilde{\chi}^{\pm}_{1,2}}$
                 &286, 19330 & 493,6021  &664,4542 &1118,6275 \\
\hline $m_{ \tilde{u}_{L,R}}$
                 &21389,21132   & 20230,20115   &18653,18574 &26187,26079  \\
$m_{\tilde{t}_{1,2}}$
                 &7389,8175  & 3465,5356    &3089,5447 &4376,7901 \\
\hline $m_{ \tilde{d}_{L,R}}$
                 &21389,21513 & 20230,20333  &18653,18742 &26187,26304   \\
$m_{\tilde{b}_{1,2}}$
                 &7836,8234 & 5417,6047    &5534,6584 &8038,9652   \\
\hline
$m_{\tilde{\nu}_{1}}$
                 &  21196 & 20128  &18565 &26037       \\
$m_{\tilde{\nu}_{3}}$
                 & 15502 & 15066   &14032 &19441     \\
\hline
\hline
$m_{ \tilde{e}_{L,R}}$
                &21193,21717 & 20123,20416   &18559,18779 & 26027,26319\\
$m_{\tilde{\tau}_{1,2}}$
                &7490,15463 & 8048,15079    &7796,14042 &9984,19455\\
\hline

$\Omega_{CDM}h^{2}$
                &12642    &190 &972 & 1377\\
\hline
$R_{tb\tau}$    &  1.06 & 1.00  & 1.05 & 1.07\\
\hline
$BF(\tg\to b\bar{b}\tz_i )$ & 0.33 & 0.13 & 0.07 & 0.06  \\
$BF(\tg\to t\bar{t}\tz_i )$ & 0.15  & 0.15 & 0.69 & 0.75   \\
$BF(\tg\to t\bar{b}\twm_j +c.c. )$ & 0.45 & 0.33 & 0.22 & 0.18 \\
\hline
\hline
\end{tabular}
\caption{Sparticle and Higgs masses (in GeV).  
All of these benchmark points satisfy the various constraints
mentioned in Section~\ref{sec:scan} and are compatible with Yukawa unification.
Point 1  exhibits a solution near the current reach limit of LHC.
Point 2 exhibits `perfect' Yukawa unification. 
Point 3 displays an example of a relatively heavy gluino within reach of LHC14. 
Point 4 represents a solution with the heaviest gluino ($\sim 3\,{\rm TeV}$) we have in our scans;
it is likely beyond reach of LHC. The uncertainty in the Higgs mass ($m_h$) estimates is about $\pm 2$ GeV.
\label{tab:bm}}
\end{table}

%%%%%%%%%%%%%%%%%%%%%%%%%%%%%%%%%%%%%%%%%%%

In addition, $\tw_1\tz_2$ production can occur at large rates for the Yukawa-unified BM points\cite{yu_lhc7}.
For all cases listed, $\tw_1\to\tz_1 W$ at $\sim 100\%$ branching fraction, and $\tz_2\to\tz_1 h$
with typically a branching fraction $\agt 90\%$. Recently, it has been pointed out\cite{wh} that
the process $pp\to\tw_1\tz_2\to Wh\tz_1\tz_1$ should be visible at LHC with $\sqrt{s}=14$ TeV and 
$\sim 100-1000$ fb$^{-1}$ of integrated luminosity. This latter gaugino pair production signal
offers a second corroborative channel for claiming a SUSY discovery in models with
lighter gauginos and decoupled squarks and sleptons. In addition, in the $Wh$ channel, the
$p_{T}(h)$ distribution may allow a chargino/neutralino mass extraction provided a very 
large data sample is acquired. Likewise, for $m_{\tg}\sim 500-800$ GeV, then $\tz_2\to\tz_1 Z$. In this case, 
$\tw_1\tz_2$ production will yield $WZ+MET$ events, for which the $WZ\to 3\ell$ and possibly 
$WZ\to \ell^+\ell^-+jets$ signatures may be visible at LHC7\cite{wz}.

For point 2 in Table~\ref{tab:bm}, we show a case with essentially perfect Yukawa unification, 
$\R=1.0$, but with $m_{\tg}=1375$ GeV. In this case, the combined branching fraction for gluinos
into top quark final states has increased to $\sim 86\%$. With such a heavy gluino, this case
would likely be beyond the reach of LHC7\cite{lhc7}. However, it should be within reach of LHC
with $\sqrt{s}=14$ TeV (LHC14), which should start operating around 2015. 
Benchmark point 3 in Table~\ref{tab:bm} shows a case with $m_{\tg}=1853$ GeV and decoupled scalars.
This case, with such a heavy gluino mass, lies right around the ultimate reach of LHC14 with
100 fb$^{-1}$, in a search for gluino pair production. However, the $Wh$ search channel 
from $\tw_1\tz_2$ production may be competitive with gluino pair searches in this case 
assuming 100-1000 fb$^{-1}$ of integrated luminosity.

The last point 4 in Table~\ref{tab:bm} corresponds to $\R=1.07$, but with $m_{\tg}=2991$ GeV
and decoupled scalars. This is the case with the largest gluino mass we were able to find while
requiring $\R<1.1$ and $m_h\sim 125$ GeV. This case would likely lie beyond reach of LHC14 for
any luminosity upgrade. Detection of a SUSY signal in this case would likely require a
$pp$ collider with $\sqrt{s}\sim 40-100$ TeV.
%%%%%%%%%%%%%%%%%%%%%%%%%%%%%%%%%%%%%%%%

%%%%%%%%%%%%%%%%%%%%%%%%%%%%%%%%%%%%%
\section{Conclusion}
\label{sec:conclude}

Previous papers examining $t-b-\tau$ Yukawa-unified models with gaugino mass unification and $\mu >0$ have
focused on solutions with rather light gluinos: $m_{\tg}\alt 0.5$ TeV. These models are now likely all excluded by recent or soon-to-be-released LHC SUSY searches. 
In light of these earlier results, we were motivated to examine if Yukawa-unified solutions with
heavier gluinos could exist, while also requiring $m_h\sim 125$ GeV, as is recently hinted at by Atlas and CMS.
Using dedicated MCMC scans over $SO(10)$ parameter space, 
we have found solutions with excellent Yukawa unification and $m_{\tg}$ ranging up to 1.4 TeV, well beyond current
LHC search limits. Loosening the Yukawa-unification criteria to $\R<1.1$, we even find solutions 
with $m_{\tg}$ nearly 3 TeV.

We have listed four $SO(10)$ benchmark points with $m_{\tg}$ spanning the range $0.75-2.9$ TeV.
Regarding LHC SUSY searches, we note that these heavier gluino solutions will be characterized
by gluino pair production at LHC, followed by decays to final states including four $b$-jets, 
four on-shell $W$ bosons $+MET$. The gluino pair events should be rich in multiple isolated
leptons plus $b$-jets, and the dominant SM background will likely arise from four top production.

%%%%%%%%%%%%%%%%%%%%%%%%%%%%%%%%%%
\section*{Acknowledgments}
S.R  and Q.S would like to thank Ilia Gogoladze for useful discussions. This work is supported in part by the DOE Grant No. DE-FG02-04ER41305 (HB) and DE-FG02-91ER40626 
(S.R and Q.S.). 
This work used the Extreme Science and Engineering Discovery Environment (XSEDE), 
which is supported by the National Science Foundation grant number OCI-1053575.

%%%%%%%%%%%%%%%%%%%%%%%%%%%%%%%%%


\begin{thebibliography}{99}
%%%%%%%%%%%%%%%%%%%%%%%%%%%%%%%%%
%
\bibitem{old} B. Ananthanarayan, G.~Lazarides and Q.~Shafi, 
\prd{44}{1991}{1613}; ibid \plb{300}{1993}{245};
 Q.~Shafi and B.~Ananthanarayan, Trieste HEP Cosmol.1991:233-244.


\bibitem{Anderson}
G.~Anderson {\it et al.} \prd{47}{1993}{3702}; ibid \prd{49}{1994}{3660};
V. Barger, M. Berger and P. Ohmann,   
\prd{49}{1994}{4908};
M. Carena, M. Olechowski, S. Pokorski and C. Wagner,  
\npb{426}{1994}{269};
B. Ananthanarayan, Q. Shafi and X. Wang, \prd{50}{1994}{5980};
L.~J.~Hall, R.~Rattazzi and U.~Sarid, Phys.\ Rev.\  D {\bf 50}, 7048 (1994);
R. Rattazzi and U. Sarid, \prd{53}{1996}{1553};
T.~Blazek, M.~Carena, S.~Raby and C.~Wagner, \prd{56}{1997}{6919}; 
T.~Blazek and S. Raby, \plb{392}{1997}{371};
T.~Blazek and S.~Raby, \prd{59}{1999}{095002};
T.~Blazek, S.~Raby and K.~Tobe, \prd{60}{1999}{113001}; ibid
\prd{62}{2000}{055001}; 
S. Profumo, \prd{68}{2003}{015006}; C. Pallis, \npb{678}{2004}{398};
M. Gomez, G. Lazarides and C. Pallis, 
\prd{61}{2000}{123512}; ibid \npb{638}{2002}{165}; ibid \prd{67}{2003}{097701};
U. Chattopadhyay, A. Corsetti and P. Nath, \prd{66}{2002}{035003};
M. Gomez, T. Ibrahim, P. Nath and S. Skadhauge,
\prd{72}{2005}{095008}.
%
\bibitem{pati}
J.~C.~Pati and A.~Salam,
  %``Lepton Number As The Fourth Color,''
  Phys.\ Rev.\  D {\bf 10}, 275 (1974).

%
\bibitem{hb}  H. Baer, M. Diaz, J. Ferrandis and X. Tata, 
\prd{61}{2000}{111701}; H. Baer, M. Brhlik, M. Diaz, J. Ferrandis,
P. Mercadante, P. Quintana and X. Tata, \prd{63}{2001}{015007}.
%
\bibitem{bf} H. Baer and J. Ferrandis, \prl{87}{2001}{211803}. 
%
\bibitem{abbbft}  D. Auto, H. Baer, C. Balazs, A. Belyaev, J. Ferrandis 
and X. Tata, \jhep{0306}{2003}{023}.
%
\bibitem{bdr} T. Blazek, R. Dermisek and S. Raby, \prl{88}{2002}{111804};
T. Blazek, R. Dermisek and S. Raby, \prd{65}{2002}{115004};
R. Dermisek, S. Raby, L. Roszkowski and
R. Ruiz de Austri, \jhep{0304}{2003}{037};
R. Dermisek, S. Raby, L. Roszkowski and
R. Ruiz de Austri, \jhep{0509}{2005}{029}.
%
\bibitem{alt} W. Altmannshofer, D. Guadagnoli, S. Raby and D. Straub, \plb{668}{2008}{385}. 
%
\bibitem{bartol1} I.~Gogoladze, R.~Khalid, S.~Raza and Q.~Shafi,
  %``t - b - tau Yukawa unification for mu < 0 with a sub-TeV sparticle spectrum,''
  JHEP {\bf 1012}, 055 (2010);
  %arXiv:1008.2765 [hep-ph].
  %%CITATION = ARXIV:1008.2765;%%
\bibitem{bartol2} I.~Gogoladze, R.~Khalid, S.~Raza and Q.~Shafi,
  %``Higgs and Sparticle Spectroscopy with Gauge-Yukawa Unification,''
  JHEP {\bf 1106} (2011) 117.
  %[arXiv:1102.0013 [hep-ph]].
  %%CITATION = JHEPA,1106,117;%%
%
\bibitem{wagner} J. Gainer, R. Huo and C. Wagner, arXiv:1111.3639 (2011).
%
\bibitem{yu_lhc7} H. Baer, S. Kraml, A. Lessa and S. Sekmen, \jhep{1002}{2010}{055}.
%
\bibitem{yu_lhc14} H. Baer, S. Kraml, S. Sekmen and H. Summy,
\jhep{0810}{2008}{079}.
%
\bibitem{yu_atlas} G.~Aad {\em et al.} (ATLAS Collaboration),
\plb{701}{2011}{398}.
%
\bibitem{btau} H. Baer, I. Gogoladze, A. Mustafayev, S. Raza and Q. Shafi,
arXiv:1201.4412 (2012). 
%
\bibitem{mur} R.~Rattazzi and U.~Sarid, \prd {53}{1996}{1553}; 
H. Murayama, M. Olechowski and S. Pokorski, \plb{371}{1996}{57}.
%
\bibitem{imh} J. Feng, C. Kolda and N. Polonsky, \npb{546}{1999}{3}; 
J. Bagger, J. Feng and N. Polonsky, \npb{563}{1999}{3};
J. Bagger, J. Feng, N. Polonsky and R. Zhang, \plb{473}{2000}{264};
H. Baer,P. Mercadante and X. Tata, \plb{475}{2000}{289};
H. Baer, C. Balazs, M. Brhlik, P. Mercadante, X. Tata and Y. Wang, \prd{64}{2001}{015002}.
%
\bibitem{bs} H. Baer and H. Summy, \plb{666}{2008}{5};
%
\bibitem{bkss} H. Baer, S. Kraml, S. Sekmen and H. Summy, \jhep{0803}{2008}{056};
H. Baer, M. Haider, S. Kraml,  S. Sekmen and H. Summy, JCAP{\bf 0902} (2009) 002.
%
\bibitem{lhc7} H. Baer, V. Barger, A. Lessa and X. Tata, 
\jhep{1006}{2010}{102} and arXiv:1112.3044.
%
\bibitem{atlas} G. Aad {\it et al.} (ATLAS collaboration),
arXiv:1109.6572 (2011).
%
\bibitem{cms} S. Chatrchyan {\it et al.} (CMS collaboration), 
\prl{107}{2011}{221804}.
%
\bibitem{higgslhc} F.~Gianotti (ATLAS Collaboration), talk at CERN public seminar,
Dec.~13, 2011; ATLAS collaboration, ATLAS-CONF-2011-163 (2011);
G.~Tonelli (CMS Collaboration), talk at CERN public seminar, Dec.~13, 2011.
%
%
\bibitem{isajet} ISAJET v7.80, by H. Baer, F. Paige, S. Protopopescu and
X. Tata, hep-ph:0312045.
%
\bibitem{isasugra} H. Baer, C. H. Chen, R. Munroe, F. Paige and X. Tata, 
\prd{51}{1995}{1046};
H.~Baer, J.~Ferrandis, S.~Kraml and W.~Porod, \prd{73}{2006}{015010}.
%
\bibitem{Hisano:1992jj} J.~Hisano, H.~Murayama, and T.~Yanagida,
%{\it Nucleon decay in the minimal
  %supersymmetric SU(5) grand unification},
  { Nucl.~Phys.} {\bf B402} (1993) 46.
Y.~Yamada,
%{\it SUSY and GUT threshold effects in SUSY SU(5) models},
{ Z.~Phys.} {\bf C60} (1993) 83;
 J.~L.~Chkareuli and I.~G.~Gogoladze,
  %``Unification picture in minimal supersymmetric SU(5) model with string
  %remnants,''
  Phys.\ Rev.\  D {\bf 58}, 055011 (1998).
%  [arXiv:hep-ph/9803335].
%
\bibitem{dr3} H. Baer, S. Kraml and S. Sekmen, \jhep{0909}{2009}{005}.
%
\bibitem{pbmz} D.~Pierce, J.~Bagger, K.~Matchev and R.~Zhang,
Nucl.~Phys.~{\bf B491}, 3 (1997). 
%
\bibitem{Olechowski:1994gm}
  M.~Olechowski and S.~Pokorski,
  %``Electroweak symmetry breaking with nonuniversal scalar soft terms and large
  %tan beta solutions,''
  Phys.\ Lett.\  B {\bf 344}, 201 (1995);
  %[arXiv:hep-ph/9407404].
  %%CITATION = PHLTA,B344,201;%%
  %\cite{Matalliotakis:1994ft}
%\bibitem{Matalliotakis:1994ft}
  D.~Matalliotakis and H.~P.~Nilles,
  %``Implications of nonuniversality of soft terms in supersymmetric grand
  %unified theories,''
  Nucl.\ Phys.\  B {\bf 435}, 115 (1995);
%  [arXiv:hep-ph/9407251].
  %%CITATION = NUPHA,B435,115;%%
%\cite{Murayama:1995fn}
%\bibitem{Murayama:1995fn}
  H.~Murayama, M.~Olechowski and S.~Pokorski,
  %``Viable t - b - tau Yukawa unification in SUSY SO(10),''
  Phys.\ Lett.\  B {\bf 371}, 57 (1996).
%  [arXiv:hep-ph/9510327].
  %%CITATION = PHLTA,B371,57;%%
%
\bibitem{Chamseddine:1982jx}
 A.~Chamseddine, R.~Arnowitt and P.~Nath, Phys.\ Rev.\ Lett.\ {\bf 49} (1982) 970;
R.~Barbieri, S.~Ferrara and C.~Savoy, Phys.\ Lett.\ {\bf B119}
(1982) 343; N.~Ohta, Prog.\ Theor.\ Phys.\ {\bf 70} (1983) 542;
L.~J.~Hall, J.~D.~Lykken and S.~Weinberg, Phys.\ Rev.\ {\bf D27}
(1983) 2359; for a review see  S.~Weinberg, {\it The Quantum Theory
of Fields: Volume 3, Supersymmetry,
 Cambridge University Press (2000) 442p}.
%
\bibitem{Drees:1986vd}
  M.~Drees,
  %``Intermediate Scale Symmetry Breaking and the Spectrum of Super Partners in
  %Superstring Inspired Supergravity Models,''
  Phys.\ Lett.\  B {\bf 181}, 279 (1986);
  %%CITATION = PHLTA,B181,279;%%
%\cite{Kolda:1995iw}
%\bibitem{Kolda:1995iw}
  C.~F.~Kolda and S.~P.~Martin,
  %``Low-energy supersymmetry with D term contributions to scalar masses,''
  Phys.\ Rev.\  D {\bf 53}, 3871 (1996)
  [arXiv:hep-ph/9503445].
  %%CITATION = PHRVA,D53,3871;%%
%
\bibitem{:2009ec}
    [Tevatron Electroweak Working Group and CDF Collaboration and D0 Collab],
  %``Combination of CDF and D0 Results on the Mass of the Top Quark,''
  arXiv:0903.2503 [hep-ex].
  %%CITATION = ARXIV:0903.2503;%%
%
%\cite{Belanger:2009ti}
\bibitem{Belanger:2009ti}
  G.~Belanger, F.~Boudjema, A.~Pukhov and R.~K.~Singh,
  %``Constraining the MSSM with universal gaugino masses and implication for
  %searches at the LHC,''
  JHEP {\bf 0911}, 026 (2009);
  %[arXiv:0906.5048 [hep-ph]].
  %%CITATION = JHEPA,0911,026;%%
H.~Baer, S.~Kraml, S.~Sekmen and H.~Summy,
  %``Dark matter allowed scenarios for Yukawa-unified SO(10) SUSY GUTs,''
  JHEP {\bf 0803}, 056 (2008).
  %[arXiv:0801.1831 [hep-ph]].
  %%CITATION = JHEPA,0803,056;%%
%  
   \bibitem{Nakamura:2010zzi}
  K. Nakamura {\it et al.} [ Particle Data Group Collaboration ],
  %``Review of particle physics,''
  J.\ Phys.\ G {\bf G37}, 075021 (2010).

%
\bibitem{bsg} H.~Baer and M.~Brhlik, \prd{55}{1997}{4463};
H.~Baer, M.~Brhlik, D.~Castano and X.~Tata, \prd{58}{1998}{015007};
%
\bibitem{bmm} K.~Babu and C.~Kolda, \prl{84}{2000}{228};
A.~Dedes, H.~Dreiner and U.~Nierste, \prl{87}{2001}{251804};
J.~K.~Mizukoshi, X.~Tata and Y.~Wang, \prd{66}{2002}{115003}.
%
\bibitem{mamoudi} D.~Eriksson, F.~Mahmoudi and O.~Stal, \jhep{0811}{2008}{035}.
%

%\cite{Schael:2006cr}
\bibitem{Schael:2006cr}
  S.~Schael {\it et al.}  %[ALEPH Collaboration and DELPHI Collaboration and
      %            L3 Collaboration and ],
  %``Search for neutral MSSM Higgs bosons at LEP,''
  Eur.\ Phys.\ J.\  C {\bf 47}, 547 (2006).
 % [arXiv:hep-ex/0602042].
%
\bibitem{CMS_plus_LHCb}
{\bf CMS} and {\bf LHCb} Collaborations.
``Search for the rare decay  $B_{s,d} \rightarrow  \mu^+\mu^-$ at the LHC'',
LHCb-CONF-2011-047; CMS PAS BPH-11-019.
%


\bibitem{Barberio:2007cr}
  E.~Barberio {\it et al.}  [Heavy Flavor Averaging Group (HFAG)
                  Collaboration],
  %``Averages of b-hadron properties at the end of 2006,''
  arXiv:0704.3575 [hep-ex].

\bibitem{Barberio:2008fa}
  E.~Barberio {\it et al.}  [Heavy Flavor Averaging Group],
  %``Averages of $b-$hadron and $c-$hadron Properties at the End of 2007,''
  arXiv:0808.1297 [hep-ex].
  %%CITATION = ARXIV:0808.1297;%%
%
\bibitem{h125} H. Baer, V. Barger and A. Mustafayev, arXiv:1112.3017 (2011).
%
\bibitem{wh} H. Baer, V. Barger, A. Lessa, W. Sreethawong and X. Tata, 
arXiv:1201.2949 (2012).
%
\bibitem{wz} H. Baer, V. Barger, S. Kraml, A. Lessa, W. Sreethawong and X. Tata, 
UH-511-1185-12 (2012).
%
\end{thebibliography}
\end{document}